# Positivity of Quasilocal Mass

Chiu-Chu Melissa Liu* and Shing-Tung Yau†

*Harvard University, Department of Mathematics, One Oxford Street, Cambridge, Massachusetts 02138, USA*
(Received 2 March 2003; published 10 June 2003)

Motivated by the important work of Brown and York on quasilocal energy, we propose definitions of quasilocal energy and momentum surface energy of a spacelike 2-surface with a positive intrinsic curvature in a spacetime. We show that the quasilocal energy of the boundary of a compact spacelike hypersurface which satisfies the local energy condition is strictly positive unless the spacetime is flat along the spacelike hypersurface.

*Introduction.*—There have been many attempts to define quasilocal energy in general relativity (see [1] for a historical survey and a sample of relevant literature). In [1,2], Brown and York obtained definitions of surface stress-energy-momentum density, quasilocal energy, and conserved charges from a Hamiltonian-Jacobi analysis of the gravitational action. Hawking and Horowitz gave a similar derivation in [3].

The Brown-York quasilocal energy and angular momentum are defined for a spacelike 2-surface which bounds a compact spacelike hypersurface in a time orientable spacetime. They have desirable properties such as specializing to Arnowitt-Deser-Misner (ADM) energy momentum and Bondi-Sachs energy momentum in suitable limits [1,4]. Brown and York also justified their definitions with thermodynamics of black-hole physics.

Motivated by the Brown-York work [1,2], we propose definitions of quasilocal energy and surface momentum density for spacelike 2-surfaces with positive intrinsic curvature. Our definition of quasilocal mass also arises naturally from calculations in the second author's recent work [5] on black holes as a candidate for positivity, which is an essential property for any definition of mass. (Our definition of quasilocal mass was found initially by reasons motivated by understanding the second author's calculations in [5]. However, the works of Brown and York certainly make it well motivated from the point of view of physics.)

Quasilocal mass is important as it has several potential applications. It can be used to define binding energy of stars. It can also be useful for numerical relativity as it tells us how to cut off the data on a noncompact region to a compact region where the quasilocal mass of the compact region approximates the ADM mass of the noncompact region. When we evolve the data according to Einstein equations, we need local control of energy to see how the space changes. The positivity of quasilocal mass is essential for such an investigation. This could be regarded as a generalization of the energy method in the theory of a nonlinear hyperbolic system.

*Brown and York's definitions.*—Let $\Omega$ be a compact spacelike hypersurface in a time orientable spacetime $M$. We do not assume the boundary $\partial \Omega$ is connected. Let $g_{ij}$ denote the positive definite metric on $\Omega$, and let $K_{ij}$ denote the extrinsic curvature of $\Omega$ in $M$. Let $\Sigma$ be a connected component of $\partial \Omega$, and let $k$ be the trace of the extrinsic curvature $k_{ab}$ of $\Sigma$ in $\Omega$ with respect to the outward unit normal $v^i$. This definition of $k$ is the negative of the definition in [2]. Define

$$\epsilon_1 = -\frac{1}{8\pi G} k, \qquad j_1^i = -\frac{1}{8\pi G} v_j (K g^{ij} - K^{ij}),$$

where $G$ is Newton's constant, and $K = g^{ij} K_{ij}$ is the trace of the extrinsic curvature of $\Omega$ in $M$. The three vector field $j_1^i$ along $\Sigma$ can be decomposed into components tangential and normal to $\Sigma$. The tangential component can be viewed as a vector field $j_1^a$ on $\Sigma$, and the normal component is given by $-(p/8\pi G) v^i$, where $p = K - K^{ij} v_i v_j$.

Suppose that $\Sigma$ has positive intrinsic curvature so that $\Sigma$ is topologically a 2-sphere. By Weyl's embedding theorem, $\Sigma$ can be isometrically embedded into the Euclidean three space $\mathbb{R}^3$ such that the extrinsic curvature $(k_0)_{ab}$ is positive definite. Let $\bar{\Omega}$ be the region in $\mathbb{R}^3$ enclosed by $\Sigma$. The embedding $\bar{\Omega} \subset \mathbb{R}^3 \subset \mathbb{R}^{3,1}$ gives $\epsilon_0 = -k_0/8\pi G$ and $j_0^i = 0$, where $\mathbb{R}^{3,1}$ is the Minkowski spacetime. The isometric embedding $\Sigma \subset \mathbb{R}^3$ is unique up to isometry of $\mathbb{R}^3$, so $k_0$ is determined by the metric of $\Sigma$.

The energy surface density $\epsilon$ and the momentum surface density $j^a$ are defined by

$$\epsilon = \epsilon_1 - \epsilon_0 = -\frac{1}{8\pi G}(k - k_0),$$
$$j^a = j_1^a - j_0^a = -\frac{1}{8\pi G} \sigma_i^a v_j (K g^{ij} - K^{ij}),$$

where $\sigma_i^a$ is the projection to the tangent space of $\Sigma$. The quasilocal energy is defined by

$$E = \int_\Sigma \epsilon = \frac{1}{8\pi G} \int_\Sigma (k_0 - k).$$

The angular momentum with respect to a Killing vector field $\phi^a$ on $\Sigma$ is defined by

$$J = \int_\Sigma j^a \phi_a = \frac{1}{8\pi G} \int_\Sigma \phi_a \sigma_i^a v_j (K^{ij} - K g^{ij}).$$

The above quasilocal energy $E$ and angular momentum $J$ depend on the spacelike hypersurface $\Omega$. Let us denote them by $E(\Sigma, \Omega)$ and $J(\Sigma, \Omega, \phi^a)$.

*Definition of quasilocal energy.*—Let $u^\nu$ denote the future timelike unit normal of $\Omega$ in the time orientable spacetime $M$, and view $v^i$ as a four vector field $v^\nu$ defined along $\Sigma$. Then $ku^\nu - pv^\nu$ is a four vector field defined along and normal to $\Sigma$. The vector $ku^\nu - pv^\nu$ can be expressed in terms of null normals. We use the notation in [6]. Let $l^\nu$, $n^\nu$ be outward and inward null normals in the sense that $l_\nu v^\nu > 0$ and $n_\nu v^\nu < 0$. Let $2\rho$ and $-2\mu$ denote the traces of the extrinsic curvature of $\Sigma$ with respect to $l^\nu$ and $n^\nu$, respectively. The product $\rho\mu$ depends only on $l_\nu n^\nu$, which we fix to be $-1$. Then

$$ku^\nu - pv^\nu = 2(\rho n^\nu + \mu l^\nu). \quad (1)$$

We assume $k > |p|$ so that $ku^\nu - pv^\nu$ is future timelike. We have $8\rho\mu = k^2 - p^2 > 0$. It is clear from (1) that, if $\tilde{\Omega}$ is another spacelike hypersurface such that $\Sigma$ is a connected component of $\partial\tilde{\Omega}$ and the corresponding four vector field $\tilde{k}\tilde{u}^\nu - \tilde{p}\tilde{v}^\nu$ is future timelike, then $\tilde{k}\tilde{u}^\nu - \tilde{p}\tilde{v}^\nu = ku^\nu - pv^\nu$.

The embedding $\Sigma \subset \mathbb{R}^3 \subset \mathbb{R}^{3,1}$ gives similarly defined quantities $\rho_0$ and $\mu_0$. We have $8\rho_0\mu_0 = k_0^2 > 0$. We define the quasilocal energy of $\Sigma$ to be

$$E(\Sigma) = \frac{1}{\sqrt{8}\pi G} \int_\Sigma (\sqrt{\rho_0 \mu_0} - \sqrt{\rho\mu}) \quad (2)$$

$$= \frac{1}{8\pi G} \int_\Sigma (k_0 - \sqrt{8\rho\mu}). \quad (3)$$

Note that $E(\Sigma, \Omega) \leq E(\Sigma)$.

The quantity $\sqrt{8\rho\mu} = \sqrt{k^2 - p^2}$ is the boost invariant mass defined by Lau in [7]. The quantity $\rho\mu$ also appears in the definition of Hawking mass [8]. Our definition of quasilocal energy should be compared with the invariant quasilocal energy (IQE) proposed by Epp in [9], where we chose the reference energy differently. (We found Epp's paper after we completed the first version of this Letter.)

*Definitions of momentum surface density and angular momentum.*—Let $j^a$ be the momentum surface density of $\Sigma$ defined by Brown and York. There exist functions $f$ and $g$ on $\Sigma$, unique up to addition of a constant, such that

$$j^a = \epsilon^{ab} f_b + g^a.$$

In the above decomposition, $g^a$ depends on the spacelike hypersurface $\Omega$, but $\bar{j}^a = \epsilon^{ab} f_b$ does not. We define $\bar{j}^a$ to be the momentum surface density of $\Sigma$. The field strength defined by Lau in [7] is given by

$$F_{ab} = \delta_a j_b - \delta_b j_a = \delta_a \bar{j}_b - \delta_b \bar{j}_a = -(\Delta f)\epsilon_{ab},$$

which is boost invariant. We have $*F = -\Delta f$.

Embed $\Sigma$ in $\mathbb{R}^3$ as before. Let $\phi^i$ be a Killing vector field on $\mathbb{R}^3$ which generates a rotation. We define the angular momentum with respect to $\phi^i$ to be

$$J(\Sigma, \phi^i) = \int_\Sigma \bar{j}_a (\sigma_0)_i^a \phi^i,$$

where $(\sigma_0)_i^a$ is the projection of $\mathbb{R}^3$ to the tangent space of $\Sigma \subset \mathbb{R}^3$.

*Positivity of quasilocal energy.*—Let $\Omega$ be a compact spacelike hypersurface in a time orientable spacetime $M$. Let $g_{ij}$ denote the metric of $\Omega$, and let $K_{ij}$ denote the extrinsic curvature of $\Omega$ in $M$, as before. The local mass density $\mu$ and the local current density $J^i$ on $\Omega$ are related to $g_{ij}$ and $K_{ij}$ by the constraint equations

$$\mu = \tfrac{1}{2}(R - K^{ij}K_{ij} + K^2), \qquad J^i = D_j(K^{ij} - Kg^{ij}),$$

where $K = g^{ij}K_{ij}$ is the trace of the extrinsic curvature, and $R$ is the scalar curvature of the metric $g_{ij}$.

Theorem 1.—*Let $\Omega$, $\mu$, $J$ be as above. We assume that $\mu$ and $J^i$ satisfy the local energy condition*

$$\mu \geq \sqrt{J^i J_i},$$

*and the boundary $\partial\Omega$ has finitely many connected components $\Sigma_1, \ldots, \Sigma_l$, each of which has positive intrinsic curvature. Then the quasilocal energy*

$$E(\partial\Omega) = \sum_{\alpha=1}^l E(\Sigma_\alpha).$$

*of $\partial\Omega$ is strictly positive unless $M$ is a flat spacetime along $\Omega$. In this case, $\partial\Omega$ is connected and will be embedded into $\mathbb{R}^3 \subset \mathbb{R}^{3,1}$ by Weyl's embedding theorem.*

When the extrinsic curvature $K_{ij}$ of $\Omega$ in $M$ vanishes, the local energy condition reduces to $R \geq 0$, and we have $E(\Sigma_\alpha, \Omega) = E(\Sigma_\alpha)$. In this case, Shi and Tam proved in [10] that $E(\Sigma_\alpha) \geq 0$ for each $\alpha$, and $E(\Sigma_\alpha) = 0$ for some $\alpha$ if and only if $\partial\Omega$ is connected and $\Omega$ is a domain in $\mathbb{R}^3$.

We now briefly describe Shi and Tam's proof. Each $\Sigma_\alpha$ can be isometrically embedded to $\mathbb{R}^3$ as a strictly convex hypersurface diffeomorphic to a 2-sphere. Let $N$ be the complete three-dimensional manifold obtained by gluing the region $E_\alpha$ exterior to $\Sigma_\alpha$ in $\mathbb{R}^3$ to $\Omega$ along $\Sigma_\alpha$. The region $E_\alpha$ is foliated by dilations $\Sigma_{\alpha,r}$ of $\Sigma_\alpha$, where $r$ is the distance from $\Sigma_\alpha$. Shi and Tam showed that the Euclidean metric on $E_\alpha$ can be deformed radially to a metric with zero scalar curvature such that its restriction to $\Sigma_\alpha$ is unchanged and the mean curvature of $\Sigma_\alpha$ in $E_\alpha$ coincides with that in $\Omega$. This gives an asymptotically flat metric on $N$ which is smooth away from $\partial\Omega$ and Lefschitz near $\partial\Omega$. Shi and Tam proved that the positive mass theorem holds for such a metric. The quasilocal energy of $\Sigma_{\alpha,r}$ is nonincreasing in $r$ and tends to the ADM mass of the end $E_\alpha$ as $r \to \infty$, from which one derives the result.

We shall reduce Theorem 1 to Shi and Tam's result by a procedure used by Schoen and the second author in [11] and also by the second author in [5].

As in [11], we consider the following equation proposed by Jang [12]:

$$\left(g^{ij} - \frac{f^i f^j}{1 + |\nabla f|^2}\right)\left(\frac{f_{ij}}{\sqrt{1 + |\nabla f|^2}} - K_{ij}\right) = 0. \quad (4)$$

We first assume that there is no apparent horizon in $\Omega$. Here an apparent horizon is a smoothly embedded 2-sphere $S$ in $\Omega$ satisfying $k_s - p_s = 0$ or $k_s + p_s = 0$, where $k_s$ is the extrinsic curvature of $S$ with respect to the inward unit normal $\nu_s^i$, and $p_s = K - K_{ij}\nu_s^i\nu_s^j$. Under this assumption, there exists a solution $f$ to (4) on $\Omega$ such that $f|_{\partial\Omega} \equiv 0$. The induced metric of the graph $\Omega_f \cong \Omega$ of $f$ in $(\Omega \times \mathbb{R}, g_{ij}dx^i dx^j + dt^2)$ is $\bar{g}_{ij} = g_{ij} + f_i f_j$. Let $\bar{R}$ be the scalar curvature of the metric $\bar{g}_{ij}$, and let $h_{ij}$ be the extrinsic curvature of $\Omega_f$ in $\mathbb{R} \times \Omega$. Then

$$\bar{R} \geq \sum (h_{ij} - K_{ij})^2 + 2\sum_i (h_{i4} - K_{i4})^2 - 2\sum_i \bar{D}_i(h_{i4} - K_{i4}), \quad (5)$$

where the index 4 corresponds to the downward unit normal to $\Omega_f$, $\bar{D}_i$ is the covariant derivative of $\bar{g}_{ij}$. The above inequality implies that there is a unique solution to

$$\bar{\Delta}u - \tfrac{1}{8}\bar{R}u = 0 \quad (6)$$

on $\Omega$ such that $u|_{\partial\Omega} = 1$. The solution is everywhere positive, so $\hat{g}_{ij} = u^4 \bar{g}_{ij}$ is a metric with zero scalar curvature and coincides with $\bar{g}_{ij}$ on $\partial\Omega$. Let $\bar{k}$ and $\hat{k}$ denote the traces of the extrinsic curvatures $\bar{k}_{ab}$ and $\hat{k}_{ab}$ of $\partial\Omega$ with respect to $\bar{g}_{ij}$ and $\hat{g}_{ij}$, respectively. Then $\hat{k} = \bar{k} + 4u_i\bar{\nu}^i$, where $\bar{\nu}^i$ is the outward unit normal of $\partial\Omega$ in $(\Omega, \bar{g}_{ij})$. Applying Stokes' theorem and using (6), (5), we have

$$\int_{\partial\Omega} \hat{k} = \int_{\partial\Omega} \bar{k} + 4\int_\Omega \left(u_i u^i + \frac{\bar{R}}{8}u^2\right)$$
$$\geq \int_{\partial\Omega}[\bar{k} - (h_{i4} - K_{i4})\bar{\nu}^i],$$

where the equality holds if and only if $\bar{R} = 0$, $\hat{g}_{ij} = \bar{g}_{ij}$, and $h_{ij} = K_{ij}$.

Let $\bar{u}^\mu$ be the downward unit normal of $\Omega_f$ in $\Omega \times \mathbb{R}$. Let $w^i$ and $\bar{\nu}^i$ be the unit outward normals of $\partial\Omega$ in $\Omega_0$ (the graph of the zero function) and $\Omega_f$, respectively. We view $w^i$ and $\bar{\nu}^i$ as four vector fields $w^\mu$ and $\bar{\nu}^\mu$ along $\partial\Omega$. It was computed in [5], Section 5, that

$$\bar{k} - (h_{i4} - K_{i4})\bar{\nu}^i = -\frac{\bar{u}_\nu w^\nu}{\bar{\nu}_\nu w^\nu}p + \frac{1}{\bar{\nu}_\nu w^\nu}k. \quad (7)$$

Using $(\bar{u}_\nu w^\nu)^2 + (\bar{\nu}_\nu w^\nu)^2 = w_\nu w^\nu = 1$, one can check that the right-hand side of (7) is greater or equal to $\sqrt{k^2 - p^2}$. Hence,

$$\int_{\partial\Omega}[\bar{k} - (h_{i4} - K_{i4})\bar{\nu}^i] \geq \int_{\partial\Omega}\sqrt{k^2 - p^2} = \int_{\partial\Omega}\sqrt{8\rho\mu}.$$

By Shi and Tam's result,

$$\int_{\partial\Omega} \hat{k} \leq \int_{\partial\Omega} k_0,$$

where the equality holds if and only if $\partial\Omega$ is connected and $(\Omega, \hat{g}_{ij})$ is a domain in $\mathbb{R}^3$. We conclude that

$$E(\partial\Omega) = \frac{1}{8\pi G}\int_{\partial\Omega}(k_0 - \sqrt{8\rho\mu}) \geq 0,$$

and the equality holds if and only if $\Omega$ is diffeomorphic to a domain $\Omega_0 \subset \mathbb{R}^3$ and can be isometrically embedded in $\mathbb{R}^{3,1}$ as a graph $\{[x, f(x)] \mid x \in \Omega_0\} \subset \mathbb{R}^{3,1}$ with extrinsic curvature $K_{ij}$, where $f$ is a smooth function on $\Omega_0$ which vanishes on $\partial\Omega_0$. This completes the proof of Theorem 1 if there is no apparent horizon in $\Omega$.

In general, solutions to (4) which vanish on $\partial\Omega$ are defined only on the exterior of a finite family of apparent horizons, but the above proof can be modified as in [11] to give Theorem 1.

*Quasilocal mass of the black-hole.*—When we are on the apparent horizon, the second term in (3) vanishes; only the reference extrinsic curvature $k_0$ comes in. In this case, a well-known Minkowski inequality says that

$$\int_S k_0 \geq \sqrt{16\pi A},$$

for a convex surface $S$ in the Euclidean space $\mathbb{R}^3$, where $k_0$ is the trace of the extrinsic curvature of $S$ in $\mathbb{R}^3$, and $A$ is the area of $S$. Hence, our quasilocal mass must be greater than or equal to $\sqrt{A/(4\pi G^2)}$.

Based on the second author's work in [5], we conjecture that, when the quasilocal mass of a surface with positive curvature is greater than a universal constant times the square root of its area, a black hole must form in its vicinity. The proof of such a statement will then give qualitatively a necessary and sufficient condition for a black hole to form.

*Remark.*—Booth and Mann showed in [13] that, without the assumption that the timelike boundary is orthogonal to the foliation of the spacetime, the Brown-York derivation yields boost invariant quantities.

We wish to thank R. Bousso, L. Motl, S. F. Ross, and A. Strominger for helpful conversations. We also wish to thank I. S. Booth and R. B. Mann. The second author is supported in part by the National Science Foundation under Grant No. DMS-9803347.

---

*Electronic address: ccliu@math.harvard.edu
†Electronic address: yau@math.harvard.edu